\documentclass[10pt]{article}

\usepackage{hyperref}
\usepackage{amsmath}
\usepackage{subfigure}
\usepackage{epsfig}

\begin{document}

\title{\Large \textbf{An electromagnetic black hole made of metamaterials}}

\author
{Qiang Cheng, Tie Jun Cui\footnote{tjcui@seu.edu.cn}, Wei Xiang Jiang and Ben Geng Cai\\
\small\it{State Key Laboratory of Millimeter Waves, Department of Radio Engineering}\\
\small\it{Southeast University, Nanjing 210096, China.}}


\date{}

\maketitle

Traditionally, a black hole is a region of space with huge
gravitational field, which absorbs everything hitting it. In
history, the black hole was first discussed by Laplace under the
Newton mechanics, whose event horizon radius is the same as the
Schwarzschild's solution of the Einstein's vacuum field equations.
If all those objects having such an event horizon radius but
different gravitational fields are called as black holes, then one
can simulate certain properties of the black holes using
electromagnetic fields and metamaterials due to the similar
propagation behaviours of electromagnetic waves in curved space and
in inhomogeneous metamaterials. In a recent theoretical work by
Narimanov and Kildishev, an optical black hole has been proposed
based on metamaterials, in which the theoretical analysis and
numerical simulations showed that all electromagnetic waves hitting
it are trapped and absorbed. Here we report the first experimental
demonstration of such an electromagnetic black hole in the microwave
frequencies. The proposed black hole is composed of non-resonant and
resonant metamaterial structures, which can trap and absorb
electromagnetic waves coming from all directions spirally inwards
without any reflections due to the local control of electromagnetic
fields and the event horizon corresponding to the device boundary.
It is shown that the absorption rate can reach 99\% in the microwave
frequencies. We expect that the electromagnetic black hole could be
used as the thermal emitting source and to harvest the solar light.

\newpage

The concept of black hole was first introduced by Laplace under the
Newton mechanics but has been widely used in the general relativity,
in which a Schwarzschild black hole possesses an event horizon
radius \cite{1}. Everything hitting it including the light cannot
escape from the region because the gravity potential is very
powerful, and incident lights which are tangential to this circular
horizon from outside would form a circle. Similarly, any object
whose possible potential is so powerful that it forms an event
horizon could be called as a black hole, just like a recent
theoretical work reported in Ref. \cite{3}. Actually, the absorption
properties of such a black hole are similar to those of so called
\textbf{black body} in thermodynamics \cite{4}. All these black
holes have a one-way surface, which absorbs all lights/particles
into the horizon surface without any reflections, and hence nothing
comes out from the black holes.

There are two ways to mimic the phenomena of black holes based on
electromagnetic waves and metamaterials \cite{2}. For the black hole
in general relativity, the presence of matter-energy densities
results in the motion of matter propagating in a curved spacetime
\cite{1}. Such a behaviour is very similar to the light or
electromagnetic-wave propagation in a curved space or in an
inhomogeneous metamaterial. Hence one could use the electromagnetic
waves and metamaterials to mimic the celestial mechanics by
comparing the refractive index to the metric of gravity \cite{5}.
Another way is to consider an analogy between the mechanics and
optics \cite{3}, which is revealed by the least-action principle for
the motion of particle in mechanics and the Fermat principle for the
light propagation in optics \cite{21}. This analogy generates an
optical black hole \cite{3}, or an electromagnetic black hole, whose
event horizon radius is just corresponding to the boundary of the
device. Although the electromagnetic black is not the
Schwarzschild's, it does have as good absorption behaviors as the
black body, which can absorb the light waves in all directions and
in a broad frequency band with high efficiency \cite{3}. In either
ways, however, the optical/photonic black holes based on
metamaterials were only considered in theory and numerical
simulations \cite{3,5}, and experimental verifications have not been
reported.

During the past ten years, metamaterial has been a hot research
topic in the scientific community. The advances of artificial
metamaterials in theories and experiments have offered scientists
potent ways to tailor the properties of electromagnetic waves in the
curvilinear space. Metamaterials have manifested a lot of exciting
effects and devices, such as the negative refraction,
electromagnetic invisibility cloaks, super-resolution imaging,
electromagnetic concentrators, and light trapping
\cite{6}-\cite{20}, in which the required constitutive parameters
could be fulfilled by periodic/non-periodic arrays of electric or
magnetic resonant/non-resonant particles. The current technologies
for designing and fabricating metamaterials have enabled people to
realize such functional devices with unusual electromagnetic
properties.

In this work, we realize the electromagnetic black hole in the
microwave frequencies based on the theoretical prediction to the
optical black hole using non-magnetic metamaterials \cite{3}. We
have designed and fabricated the electromagnetic black hole using
non-resonant and resonant metamaterial structures, and measured the
internal electric fields using a planar-waveguide near-field
scanning apparatus. Experimental results have good agreements to the
full-wave numerical simulations, which show obvious phenomena of
microwave bending and trapping spirally into the black hole and not
coming back. The electromagnetic black hole can absorb
electromagnetic waves coming from all directions efficiently with an
absorption rate of 99\%, which could find wide applications in the
thermal emitting and the solar-light harvesting.

In the analytical mechanics, the motion of matter attracted by the
gravitational field could be described by the Hamilton equations:
\begin{equation}
\frac{d}{dt}p(t)=-\frac{\partial{\cal H}}{\partial q}, \hskip 6mm
\frac{d}{dt}q(t)=\frac{\partial{\cal H}}{\partial p},
\end{equation}
in which $p$ is the generalized momentum, $q$ is the generalized
coordinate, and ${\cal H}$ is the Hamiltonian. The Hamiltonian
represents energy of the system, which is the sum of kinetic and
potential energy, denoted by $T$ and $V$, respectively, as ${\cal
H}=T+V, \hskip 1mm T=p^2/(2m), \hskip 1mm V=V(q)$, in which $m$ is
the mass of matter.

In the geometrical optics, the Fermat principle determines the
propagation of light. Considering that the eikonal function
$S_t(r,\theta)$ in a two-dimensional (2D) cylindrical coordinate is
expanded in series $S_t(r,\theta)=S(r,\theta)-\omega t$ as an
Hamilton-Jacobi equation, we could get the Hamiltonian in optics as
\cite{21}
\begin{equation}
{\cal
H}=\frac{\omega^2}{2\mu_0}=\frac{{p_r}^2}{2\epsilon(r)}+\frac{p_{\theta}^2}{2r^2\epsilon(r)},
\end{equation}
in which $p_r={\partial S}/{\partial r}$ and $p_{\theta}={\partial
S}/{\partial \theta}$ are the radial and angular momentum in the
cylindrical coordinate, $\omega$ is the radian frequency, $\mu_0$ is
the magnetic permeability in free space, and $\epsilon(r)$ is the
electric permittivity in the isotropic non-magnetic medium. Since
the frequency is invariant in a time-harmonic system, the above
equation reveals an energy conservation in mechanics. From the
Hamilton equation, the trajectory of geometrical optics looks like a
unit particle in a central potential which is given by \cite{3}:
$V_{\rm eff}(r)={\omega}^2c^2[\epsilon_b-\epsilon(r)]/2$, in which
$\epsilon_b$ is the background permittivity, and $c$ is the light
speed. When the inhomogeneous electric permittivity is chosen as
\begin{equation}
\epsilon(r)=\left\{
\begin{array}{ll}
\epsilon_b, & r>R \\
\rule{0in}{3ex}
\epsilon_s(r)=\epsilon_b(R/r)^2, \hskip 7mm & R_{c}\leq r \leq R \\
\rule{0in}{3ex}
\epsilon_c+i\gamma, & r<R_c\\
\end{array}
\right.,
\end{equation}
the Hamilton equation represents a Kepler problem. The above
permittivity distribution describes a layered dielectric cylinder,
which includes a lossy circular inner core and a lossless circular
shell with radially-varied permittivity, as shown in Fig. 1(a).
Here, $R_c$ and $R$ stand for the radii of inner core and outer
shell of the cylinder, while $\epsilon_b$ and $\epsilon_c$ represent
the dielectric constants of the background medium and the lossy
material inside the core. The radius of core is closely related to
the ratio of two dielectric constants:
$R_{c}=R\sqrt{\epsilon_{b}/\epsilon_{c}}$. Under the ray
approximation, it has been proved that the light hitting the
cylinder will bend spirally in the shell region, and be trapped and
absorbed by the lossy core, as illustrated in Fig. 1(a). It is also
shown that the scattering cross section per unit length of the
cylinder is nearly zero, which is independent of the polarization
state of incident waves \cite{3}. Hence the dielectric cylinder
behaves like a 2D \textbf{black hole} and the boundary represents
the event horizon, which could absorb all lights hitting it from
every direction.

When the operating frequency falls into the microwave band, the ray
approximation produces a certain error, which results in small
scattering cross sections from the black hole. Hence we expect that
the electromagnetic black hole will absorb most electromagnetic
waves hitting it from every direction with tiny scattering. Next we
will validate the electromagnetic black hole through numerical
simulations and experiments. From Eq. (3), the permittivity
distribution of the black hole's outer shell ($R_c<r<R$) varies
gradually from the inner core to the background medium. Hence the
outer shell can be realized by gradient refraction index
metamaterials, which have been used in the design of some new
concept devices such as the invisibility cloak \cite{10} and ground
cloaks \cite{12,13}.

A number of non-resonant metallic structures could be utilized as
the basic element of gradient refraction index metamaterials, such
as the circular ring, I-shaped structure, and Jerusalem cross. The
dimensions of unit geometries could be adjusted to meet the demand
of refraction indices at specific positions, to achieve the gradient
distribution. In non-resonant metamaterials, the resonant
frequencies of unit geometries are much higher than the operating
frequency, and the dispersion curves for effective permittivity and
permeability change slowly in a broad band. Considering the
anisotropy of most metamaterial unit, the electromagnetic
constitutive parameters for the black hole are adjusted for the
transverse-electric (TE) polarization in the cylindrical coordinate
as $\epsilon_{z}=\epsilon(r)$, $\mu_{\phi}=1$, and $\mu_{r}=1$. In
our design, we choose the non-resonant I-shaped structure \cite{12}
as the basic unit for the outer shell of the electromagnetic black
hole, and the electric-field-coupled (ELC) resonator \cite{19} as
the basic unit for the inner core, which has large permittivity and
large loss tangent simultaneously near the resonant frequency.

The photograph of the fabricated electromagnetic black hole is shown
in Fig. 1(b), in which the I-shaped unit cell and ELC resonator are
illustrated in Figs. 2(a) and 2(b), respectively. The black hole is
placed in the air, hence the permittivity of background medium is
simply $\epsilon_{b}=1$. In order to better demonstrate the
absorption effect, a relatively high microwave frequency ($f=18$
GHz) is selected in simulations and experiments. The sizes of both
I-shaped unit cell and ELC resonator are set as 1.8 mm, nearly
$1/10$ of the free-space wavelength. The whole black hole is
composed of 60 concentric layers, and each layer is a thin printed
circuit board (F4B, $\epsilon=2.65+i0.001$) etched with a number of
sub-wavelength unit structures. From Eq. (3), the permittivity
changes radially in the shell of black hole, and hence the unit
cells are identical in each layer but have different sizes in
adjacent layers. Since the permittivity is a constant in the lossy
core, the ELC resonators are identical in the whole region.

Figure 2 demonstrates the effective medium parameters of the
I-shaped and ELC units for the designed black hole. The full-wave
numerical tool (Microwave Studio, CST2006b) has been used to
simulate the electromagnetic properties. Following the standard
retrieval procedure \cite{22}, the effective permittivity and
permeability are obtained from the scattering parameters with the
change of geometry dimensions. To determine the relation between
geometry and medium parameters, an interpolation algorithm has been
developed to generate the final layout according to the permittivity
distribution required by the black hole. From Fig. 2(a), by changing
the height of I-shaped unit, the real part of permittivity
Re($\epsilon_z$) ranges from 1.27 to 12.64 at 18 GHz, while the
permeability components, Re($\mu_r$) and Re($\mu_{\phi}$), are
always close to unity. The imaginary parts of permittivity and
permeability for the I-shaped unit, not shown here, are small enough
to be neglected in the design of black hole. From Fig. 2(b), the
operating frequency is close to the resonant frequency of the ELC
structure, which results in a very lossy permittivity
$\epsilon_z=9.20+i2.65$ and lossy permeability components
$\mu_r=0.68-i0.01$ and $\mu_{\phi}=0.84-i0.14$ at 18 GHz.

In our design, the space between adjacent layers is 1.8 mm, hence
the radii of the black hole and the lossy core are determined
immediately as $R=108$ mm and $R_c=36$ mm. There are totally 40
layers of I-shaped structures and 20 layers of ELC structures. In
order to fix the 60 layers with different radii together, a
0.8-mm-thick styrofoam board has been used with 60 concentric
circular slots carved by the LPKF milling machine (LKPF s100). Each
layer has three unit cells in the vertical direction, and hence the
height of black hole is 5.4 mm. To investigate the interactions
between the fabricated black hole and incident TE-polarized
electromagnetic waves, a parallel-plate waveguide near-field
scanning system is used to map the field distributions near the
black hole at 18 GHz. A similar apparatus has been discussed in Ref.
\cite{23}. The separation between two plates is set as 6.5 mm, which
is larger than the height of black hole to avoid the unnecessary
dragging during measurements. The cutoff frequency of the waveguide
system for the dominant TEM mode is 23 GHz. The bottom plate is
mounted on a step motor which can translate in two dimensions. A
monopole probe is fixed inside the waveguide as the feeding source,
and a corner reflector is placed on the back of source to produce
the desired narrow beam. Four detection probes are placed on the top
plate to measure the field distributions on a plane above the black
hole under test, and each probe can scan a region of 200 mm by 200
mm independently. Hence the total scanning region is 400 mm by 400
mm with a step resolution of 0.5 mm. All probes are connected to a
microwave switch which controls the measurement sequence after each
movement of the step motors below the bottom plate. The two ports of
a vector network analyzer (Agilent PNA-L N5230C) are connected to
the feeding probe and the microwave switch respectively via cables.
Then the measurement data are sent to the controlling computer for
post processing.

To demonstrate the performance of black hole in the microwave
frequency, we first consider the case of Gaussian-beam incidence.
Figures 3(a) and (b) illustrate the distributions of simulated
electric fields $|E_z|$ at 18 GHz when a Gaussian beam is incident
on the black hole on-center and off-center, respectively. We note
that all on-center rays are directly attracted by the black hole
without reflections, and nearly all off-center rays bend in the
shell region spirally and are trapped by the core. To evaluate the
absorption of Gaussian beam by the electromagnetic black hole, we
define an absorbing rate from the Poynting theorem as
\begin{eqnarray}
\eta=P_{\rm absorb}/P_{\rm in}, \ \ P_{\rm
absorb}=-\frac{1}{2}\mbox{Re}(\oint_{s}\vec{E}\times
\vec{H}^{*}\cdot d\vec{S}),
\end{eqnarray}
in which $P_{\rm absorb}$ is the net power entering the black-hole
surface, and $P_{\rm in}$ is the incident power. Under the on-center
and off-center incidences shown in Figs. 3(a) and 3(b), the
absorbing rates are calculated as 99.94\% and 98.72\%, respectively.
Nearly all incident powers are absorbed by the microwave black hole.

In the microwave frequency, however, it is difficult to generate the
Gaussian beam. Hence we use monopole probe with corner reflector to
produce the narrow beam in our experiments, as demonstrated in Figs.
3(c)-(f). Comparing with the Gaussian beam, the produced beam is
divergent while propagating. Similar to Figs. 3(a) and 3(b), we
measure two cases when the electromagnetic waves are incident
vertically and obliquely to the black hole. As a comparison, the
full-wave numerical simulations are also given to illustrate the
absorption effect. Figures 3(c) and 3(e) illustrate the
distributions of simulated and measured electric fields $|E_z|$ at
18 GHz for the vertical-incidence case, which agree to each other
very well. It is clear that the incident beam becomes convergent
inside the shell region and then enters the lossy core of black hole
instead of being divergent in the free-space radiation. The
absorbing rate corresponding to Fig. 3(c) is 99.55\%, nearly full
absorption. When the beam is incident to the black hole at an
oblique angle $25^{\circ}$, the waves are bent toward the central
area and travel around the shell spirally with distinct absorption,
as shown in Figs. 3(d) and 3(f). Again, the simulation and
experiment results have good agreements. From Figure 3, the black
hole is a good attractor and absorber to microwaves.

Comparing the experiment and simulation results with the theoretical
prediction in the optical frequency \cite{3}, it is clear that the
design for optical black hole still holds in the microwave region.
Hence we realize the electromagnetic black hole experimentally in
the microwave frequency, which could be used to collect microwaves
and energies in free space. When the incident waves are not narrow
beams, they can also be absorbed efficiently by the black hole.
Figures 4(a) and 4(b) demonstrate the field and power distributions
inside and outside the black hole under the incidence of plane waves
(such as solar light). Obviously, nearly all incident waves hitting
the black hole are trapped to the center and do not come out with
the absorbing rate of 99.12\%. Being absorbed by the black hole, the
incident waves hitting it cannot go through, and the total fields
are nearly zero in the front region, \textbf{making a complete
shadow}. We also notice that the black hole nearly does not disturb
the electromagnetic waves in other regions. When the incident waves
are excited by a nearby monopole, the simulated and measured field
distributions are shown in Figs. 4(c) and 4(d), respectively. Like
the case of plane-wave incidence, nearly all incident waves hitting
the black hole (see the black dashed lines) are absorbed, producing
a shadow region. The simulated and measured results have good
agreements.

In summary, we have designed, fabricated, and measured an
electromagnetic black hole at the microwave frequency using
non-resonant I-shaped metamaterials and resonant ELC metamaterials
based on the theoretical study \cite{3}. We have observed that
nearly all incident waves and energies hitting the designed black
hole are attracted and absorbed. The good consistence between
theoretical and experimental results have shown the excellent
ability for metamaterials as the candidate to construct the
artificial black holes. Although the proposed electromagnetic black
is not the Schwarzschild black hole, it does have as good absorption
behaviors as the black body does in thermodynamics. Since the lossy
core can transfer the electromagnetic energies into heat energies,
we expect that the electromagnetic black hole could find important
applications in the thermal emitting and the solar-light harvesting.

This work was supported in part by the National Science Foundation
of China under Grant Nos. 60990320, 60990324, 60671015, 60496317,
and 60901011, in part by the Natural Science Foundation of Jiangsu
Province under Grant No. BK2008031, and in part by the 111 Project
under Grant No. 111-2-05.


\newpage

\section*{{{\bf List of Figure Captions}}}

\noindent \textbf{Fig. 1:} {(color online) (a) A model of black hole
composed of a gradient-index metamaterial shell and a lossy
dielectric core. (b) Photograph of the fabricated artificial black
hole based on metamaterials, which is composed of 60 concentric
layers, with ELC structures in the core layers and I-shaped
structures in the shell layers. }

\vskip 5mm

\noindent \textbf{Fig. 2:} {(color online) Effective medium
parameters for unit cells of the artificial black hole. (a) The
relation between the effective permittivity (real part of
$\epsilon_z$) and permeability (real parts of $\mu_r$ and
$\mu_{\phi}$) and the geometry dimension $m$ for the I-shaped unit.
The inset shows the sketch of the I-shaped unit, with $w=0.15$ mm
and $q=1.1m$. (b) The effective permittivity (real and imaginary
parts of $\epsilon_z$) and permeability (real parts of $\mu_r$ and
$\mu_{\phi}$) versus to the frequency for the ELC resonator. The
inset shows the sketch of the ELC unit, where $t=1.6$ mm, $g=0.3$
mm, $p=0.15$ mm, and $s=0.65$ mm.}

\vskip 5mm

\noindent \textbf{Fig. 3:} {(color online) Distributions of electric
fields $|E_z|$ for the designed black hole at the frequency of 18
GHz. An electric monopole is placed inside a corner reflector to
produce the desired incident beam with finite width. The two circles
stand for boundaries of the outer shell and the inner core. (a) The
full-wave simulation result under the on-center incidence of a
Gaussian beam. (b) The full-wave simulation result under the
off-center incidence of a Gaussian beam. (c) The full-wave
simulation result under the vertical incidence of the produced
narrow beam. (d) The full-wave simulation result under the oblique
incidence of the produced narrow beam. (e) The experimental result
under the vertical incidence of the produced narrow beam. (f) The
experimental result under the oblique incidence of the produced
narrow beam.}

\vskip 5mm

\noindent \textbf{Fig. 4:} {(color online) (a) Distributions of
electric fields for the electromagnetic black hole at the frequency
of 18 GHz when a plane wave is incident. (b) Distributions of power
flows for the electromagnetic black hole at the frequency of 18 GHz
when a plane wave is incident. (c) Full-wave simulation results of
electric fields under the excitation of a monopole at 18 GHz. (d)
Experimental results of electric fields under the excitation of the
monopole at 18 GHz.}

\newpage

\begin{figure}[htb]
 \centerline{\includegraphics[width=13.0cm]{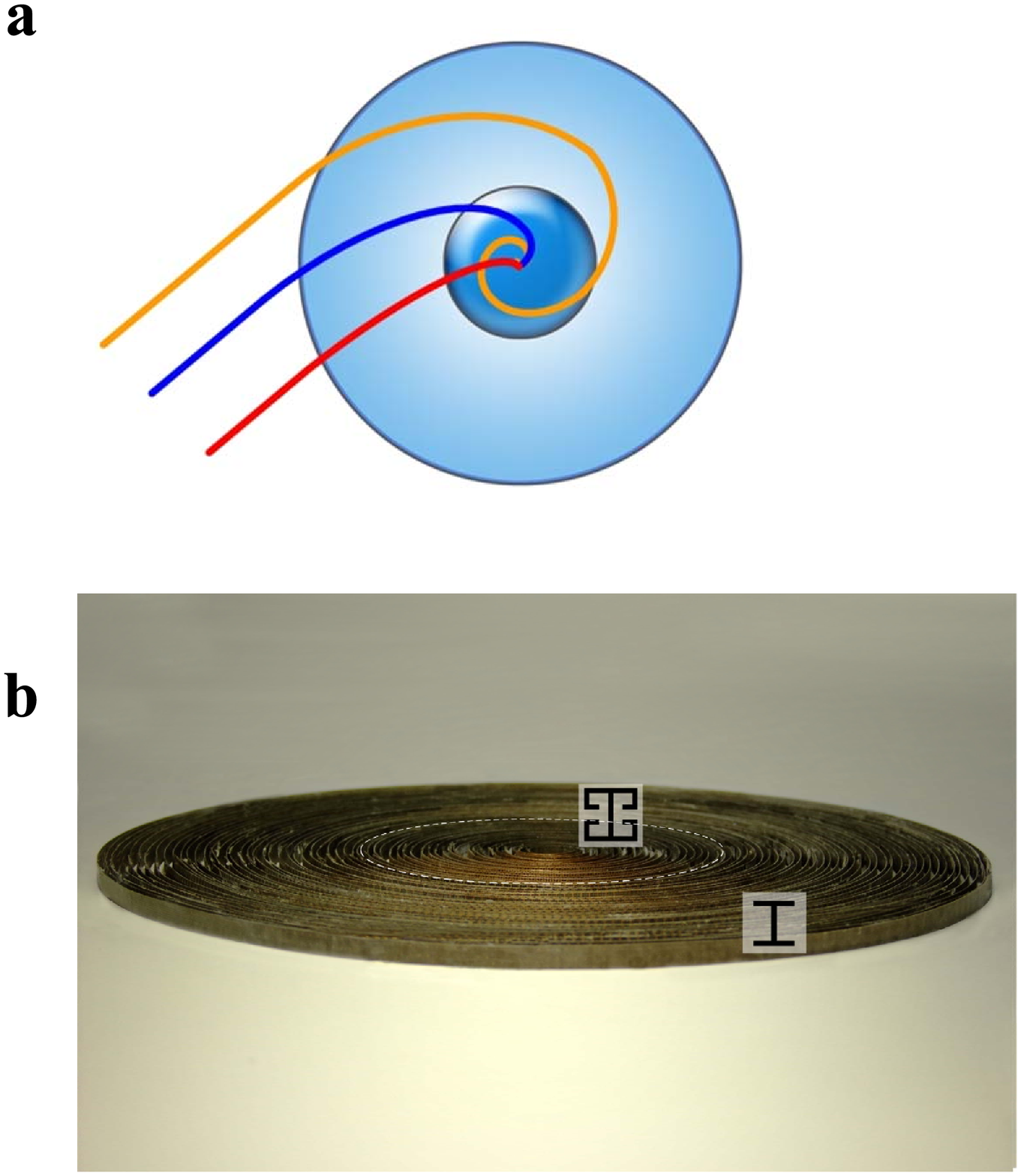}}
\caption{} \label{fig1}
\end{figure}

\begin{figure}[htb]
 \centerline{\includegraphics[width=12.0cm]{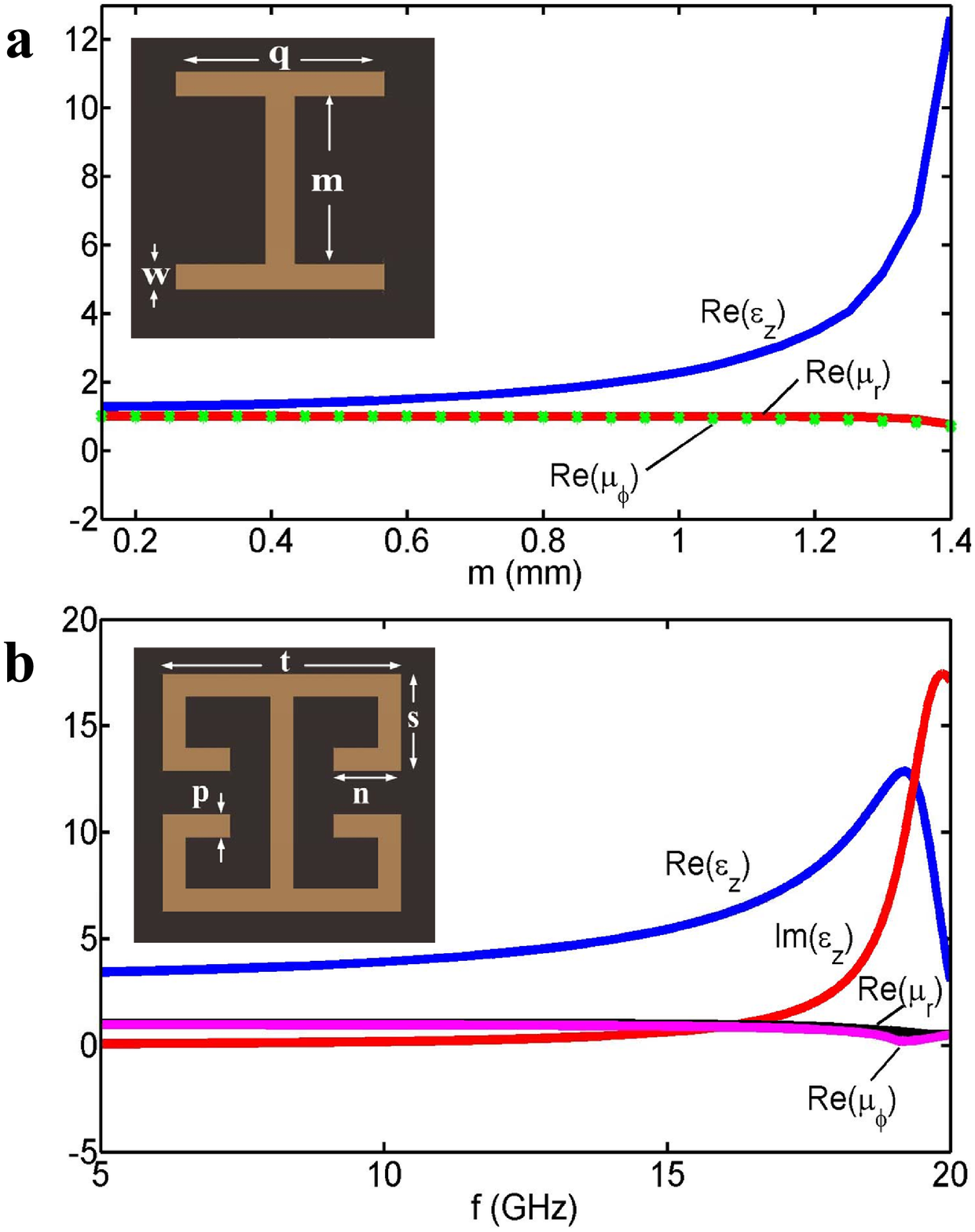}}
\caption{} \label{fig2}
\end{figure}

\begin{figure}[htb]
 \centerline{\includegraphics[width=14.0cm]{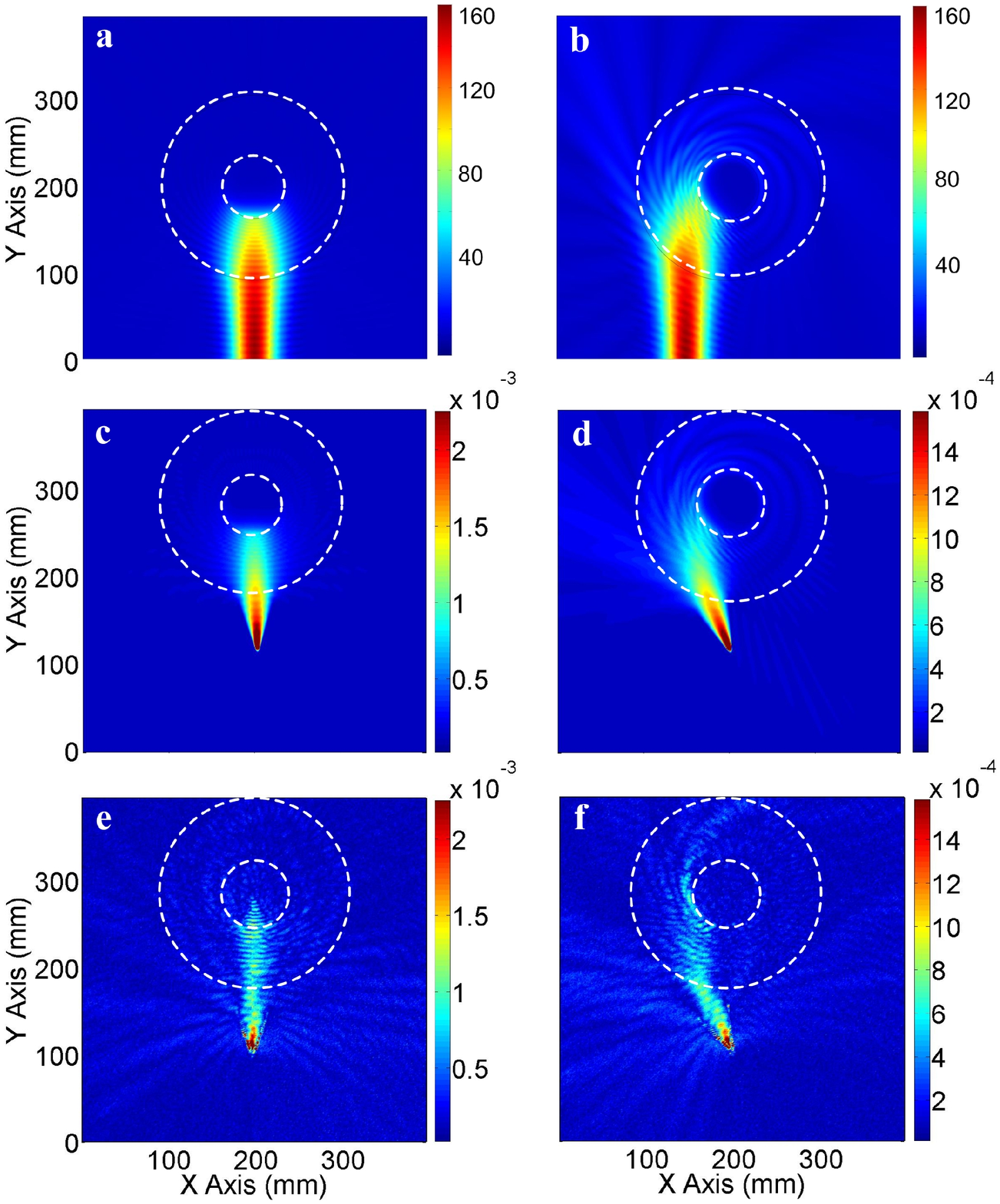}}
\caption{} \label{fig3}
\end{figure}

\begin{figure}[htb]
 \centerline{\includegraphics[width=14.0cm]{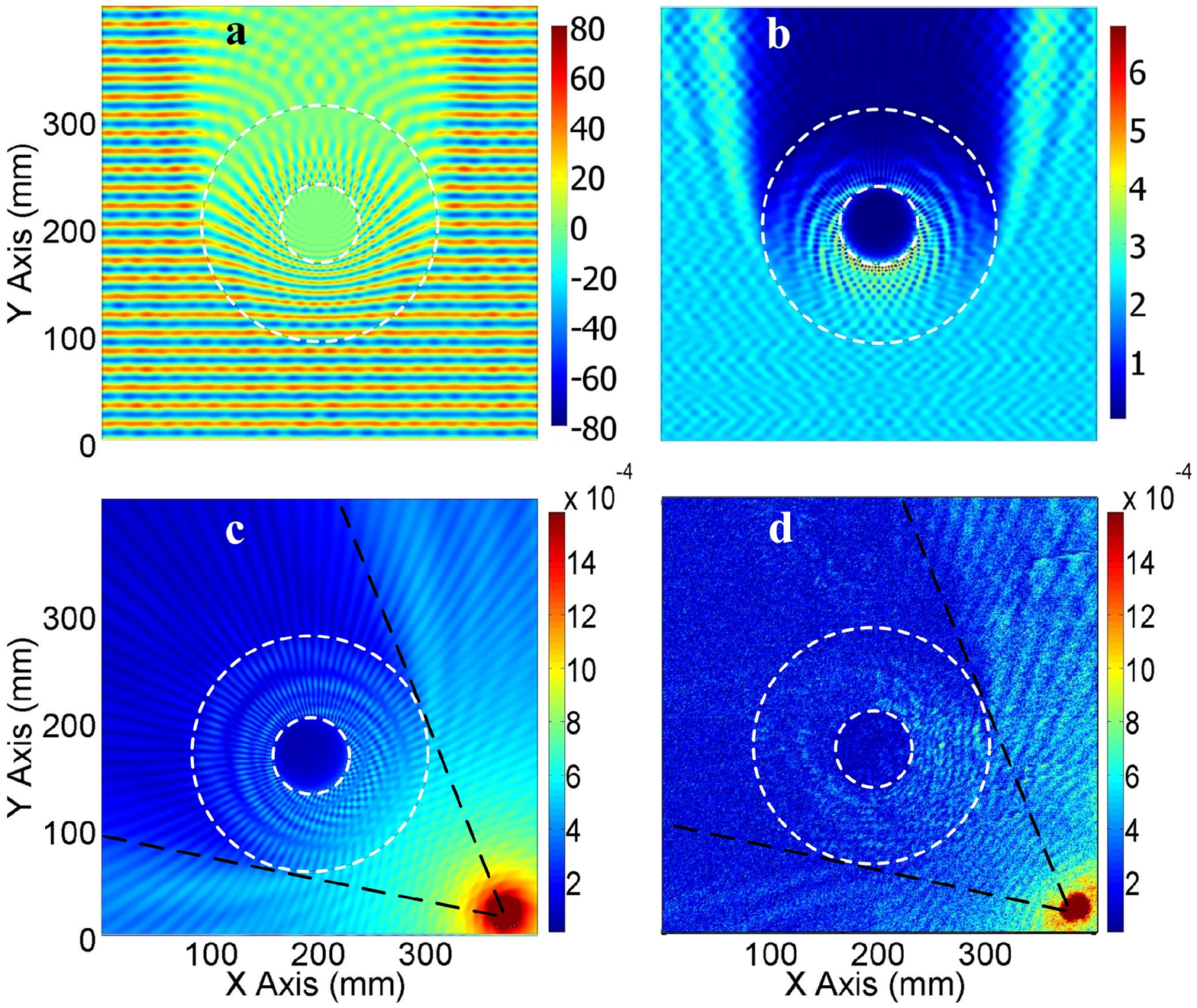}}
\caption{}
\label{fig4}
\end{figure}

\end{document}